\documentstyle[prl,aps]{revtex}
\begin{document}
\draft

\title{Anomalous Hall Effect in Double Exchange Magnets}

\author{Yong Baek Kim$^{a,b}$, Pinaki Majumdar$^b$, 
A. J. Millis$^c$, and Boris I. Shraiman$^b$}
\address{$^a$Department of Physics, The Pennsylvania State University,
University Park, PA 16802\\
$^b$Bell Laboratories, Lucent Technologies, Murray Hill, NJ 07974\\
$^c$Department of Physics and Astronomy, The Johns Hopkins University,
Baltimore, MD 21218}

\date{March 26, 1998}

\maketitle

\begin{abstract}

We investigate the possible origin of anomalous Hall effect in 
the CMR (colossal magnetoresistance) materials - the doped rare earth 
manganites - observed recently by Matl {\it et al}\cite{ong}.
It is demonstrated that the spin-orbit interaction in the double exchange 
model couples magnetization to the Berry phase associated with three
dimensional spin textures and induces a non-zero average
topological flux which in turn generates
an anomalous contribution to transverse resistivity.
The same effect, but involving the orbital Berry phase,
occurs in the model with orbital degeneracy and Coulomb repulsion. 

\end{abstract}
\pacs{PACS numbers:}

It has been known for a long time\cite{chien} that the Hall effect
in ferromagnets exhibits a low field ``anomaly" which can be 
parametrized as magnetization, $M$, dependent contribution to 
transverse resistivity: $\rho_{xy} = R_H B + R_A M$.
The anomalous Hall constant $R_A$ is usually explained in terms of
skew scattering due to spin-orbit coupling as reviewed in
Ref\cite{chien}. 
Recently  Matl {\it et al}\cite{ong} 
measured the Hall resistivity of La$_{1-x}$Ca$_x$MnO$_3$ 
with $x=0.3$
and found negative $R_A$ (in contrast with positive, hole like, $R_H$) 
which becomes 
significant above $100 K$ and peaks about $20K$ above the Curie temperature 
$T_c \approx 265K$.
This temperature dependence (which follows closely that of the 
longitudinal resistivity) contrasts with the behavior of, for example, 
Ni, where $R_A$ peaks well below $T_c$ and contradicts to
the existing theories\cite{kondo,maranzana,nozier}. 
There exist also related Hall measurements on similar material at 
relatively high temperatures\cite{jaime} and on 
La$_{1-x}$Ca$_x$CoO$_3$ ($0.1 < x < 0.5$)\cite{samoilov}. 
These new data lead us to revisit the
theory of the anomalous Hall effect.

The Manganites are described by
a model which includes double exchange, Jahn-Teller,
and electron-electron (Hubbard) interactions:
\begin{equation}
H_0 =
- t \sum_{r , \mu} c^{\dagger}_{r , \sigma , a} 
c_{r+\mu , \sigma ,a}
-J_{\rm H} \sum_{r} {\bf S}_r \cdot
c^{\dagger}_{r, \sigma ,a} {\hat \tau}_{\sigma \sigma^\prime} 
c_{r, \sigma^\prime, a}  \,+\,H_{\rm JT} \,+\,H_{e-e} \ .
\label{th}
\end{equation}
where $c^{\dagger}_{r, \sigma, a}$ is the creation operator for the
conduction electron with spin $\sigma$ and the orbital index $a = 1, 2$,
$t \sim 0.1-0.5 eV$ is the hopping matrix element
and
${\bf S}_r$ is the core spin. 
This model is suggested by the electronic structure of the
doped rare earth manganites\cite{disc,old,oldtheory} 
(Re$_{1-x}$A$_x$MnO$_3$ where Re is a rare-earth element such as 
La or Nd and A is a divalent metal ion such as Sr or Ca) where 
$1-x$ electrons occupy the orbitally degenerate $e_g$ states
($d_{x^2-y^2}$ and $d_{3z^2 - r^2}$) and couple strongly to the core $S=3/2$ 
moment via Hund's rule, $J_H \sim 1-2eV$. The on-site Coulomb repulsion
$ H_{e-e}=  U \sum_{r} n_{r, \uparrow, a} n_{r, \downarrow, a}
+ V \sum_{r, \sigma, \nu} n_{r, \nu, 1} n_{r, \sigma, 2} \ $ (where
$n_{r, \sigma, a} \equiv c^{\dagger}_{r, \sigma, a} c_{r, \sigma, a}$)
is believed to be of the same order as $J_H$\cite{varma,nagaosa}.
In addition, it has been argued\cite{millis,los} that
in the rare earth manganites an important role is played by the 
Jahn-Teller (JT) coupling of the $e_g$ electron with the phonon mode 
${\bf Q}$ describing a local tetragonal distortion which lifts 
orbital degeneracy via
$H_{\rm JT} = 
g \sum_{r } c^{\dagger}_{r, \sigma , a} {\bf Q}_r^{ab} c_{r , \sigma , b}$. 
This coupling can explain
the transition from metallic to activated transport at $T_c$ for a certain 
range of doping $x$. Below we shall discuss both the case of large
local distortions where the orbital degree of freedom is essentially
quenched and the case where 
distortions are weak and the orbital degeneracy is retained (at least on the 
scale of $k_B T$) which may be appropriate for the 
ferromagnetic metal phase of the compounds in question. 

Since the anomalous Hall effect manifests itself as the dependence of
$\rho_{xy}$ directly on the magnetization, 
it must originate from the spin-orbit interaction\cite{chien}.
It includes the (relativistic) interaction of the spin of
the itenerant electron with the local electric field and 
the magnetic interaction of the itinerant electron
with the core spins. 
The latter term,
${\bf S} \cdot {\bf p} \times {\bf r}/r^3$,
because spin-orbit coupling is quenched by the cubic crystal field
has no matrix element bewteen $e_g$ electrons
on the same site, but it does have matrix elements bewteen different sites.   
In the continuum limit, the sum of both contributions 
can be written as\cite{nozier,com2}
\begin{equation}
H_{\rm so} = i \int d^3r  \ u_{ab} \
\epsilon_{ijk} \ S_{i}(r) \ \partial_j c^{\dagger}_{ \sigma, a} (r)\
\partial_k c_{\sigma, b} (r)\ .
\label{so}
\end{equation}
As with the hopping matrix, we shall for simplicity ignore the
orbital dependence of the coupling constant and take $u_{ab} = 
\delta_{ab} u$.
With the spatially uniform spin-orbit interaction\cite{com3}
the coupling to the uniform component of
${\bf S}(r) $, {\it i.e.}, magnetization, has the form of a 
total derivative and can be integrated out to the boundary. 
Yet, we shall now show that in the presence of
interactions this {\it topological} term is 
important and generates the anomalous Hall effect.
Previous authors have studied spatially 
non-uniform spin configurations due to magnetic impurities 
or spin-wave excitations about the ferromagnetic ground state.
Here we argue that Berry phase effects caused by non-coplanar
spin configurations give the dominant contribution to $\rho_{xy}$,
at least in the limit of strong carrier-spin couplings 
appropriate to the maganites.

As a mathematical convenience to make the Berry phase physics
explicit, we use the standard Hubbard-Stratonovich approach in which
we express the interacting electron in terms of 
a non-degenerate spinless fermion $\psi_r$ and the slave boson 
fields $z_{r,\sigma}$ and $w_{r,a}$ which parametrize the spin and 
the orbital degrees of freedom respectively:
$c_{\sigma ,a} = \psi^\dagger z_\sigma w_a$.
The bosons obey the single occupancy constraint: 
$\sum_{\sigma} {\bar z}_{r,\sigma} z_{r,\sigma} =
\sum_{a} {\bar w}_{r,a} w_{r,a} = 1$ and the average fermion 
density is $\langle \psi^{\dagger}_r \psi_r \rangle = x$. 
Neglecting terms of relative order $t/J_{\rm H}$ and $t/U$,
we find
\begin{equation}
H_{\rm eff} = 
t \sum_{r, \mu } {\bf z}^{\dagger}_{r+\mu} {\bf z}_{r}
{\bf w}^{\dagger}_{r+\mu} {\bf w}_{r} \psi^{\dagger}_{r} \psi_{r+\mu}  
- J_{\rm H} \sum_{r} {\bf S}_r \cdot {\bf z}^{\dagger}_{r } {\hat {\bf \tau}} 
{\bf z}_r \psi_{r}  \psi^{\dagger}_{r}  + g
\sum_{r}  {\bf w}^{\dagger}_{r } {\bf Q}_r {\bf w}_r \,
\psi_{r}  \psi^{\dagger}_{r} \ ,
\label{tb}
\end{equation}
where ${\bf z}^{\dagger} = ({\bar z}_{\uparrow},{\bar z}_{\downarrow})$
and ${\bf w}^{\dagger} = ({\bar w}_1,{\bar w}_2)$.
The factorization of $c_{\sigma,a}$ not only enforces
single occupancy but projects onto the low energy spin and orbital state.
Assuming $J_{\rm H} \gg t$ the carrier spin and hence 
${\bf s} \equiv {\bf z}^{\dagger}_{r} {\hat {\bf \tau}} {\bf z}_r$ 
is aligned with ${\bf S}_r$. 
The generally complex spin overlap factor
${\bf z}^{\dagger}_{r+\mu} {\bf z}_{r}$ is then determined by
the Euler angles of the corresponding local moments and
is identical to the one found 
in a conventional derivation of double exchange\cite{oldtheory}.
Orbital degree of freedom results in an analogous {\it orbital} overlap factor
${\bf w}^{\dagger}_{r+\mu} {\bf w}_{r}$. Note however that since the 
tetragonal/orthorhombic phonon field ${\bf Q} = Q_1 
{\bf \tau}^x + iQ_2 {\bf \tau}^y $ is real, large $g |{\bf Q}|$ forces the
orbital ``isospin" ${\bf n} \equiv {\bf w}^\dagger {\hat \tau} {\bf w}$
into planar configurations.
We emphasize that although this formalism is the most natural one
in the strong coupling limit, we expect the physics to apply at
any coupling, with possibly different amplitude as discussed below. 

The Hamiltonian (\ref{tb}) is invariant under {\it two} local $U(1)$
transformations\cite{nagaosa,sarker}.
One of them involves the spin degree of freedom and is given by
$\psi_r \rightarrow \psi_r e^{ i \xi_r}$ and
${\bf z}_r \rightarrow {\bf z}_r e^{i \xi_r}$.
This spin $U(1)$ symmetry is broken in the ferromagnetic (FM) phase, 
but is restored in the paramagnetic (PM) phase.
The other one involves the orbital degree of freedom
and represented as
$\psi_r \rightarrow \psi_r e^{i \zeta_r}$ and
${\bf w}_r \rightarrow {\bf w}_r e^{i \zeta_r}$.
As long as the orbital ordering does not occur\cite{millis,nagaosa}
the orbital $U(1)$ symmetry is preserved.
These $U(1)$ symmetries
give rise to two gauge fields which 
are defined as
\begin{eqnarray}
a^s_{\mu} &=&  {i \over 2}
[ (\partial_{\mu} {\bf z}^{\dagger}) {\bf z} 
- {\bf z}^{\dagger} \partial_{\mu} {\bf z}] \cr
a^n_{\mu} &=&  {i \over 2}
[ (\partial_{\mu} {\bf w}^{\dagger}) {\bf w} - 
{\bf w}^{\dagger} \partial_{\mu} {\bf w} ] \ .
\end{eqnarray}
Non-trivial configurations of the gauge field are generated 
by non-coplanar arrangements of spin or orbital moments; 
they affect carrier motion just as does an external
magnetic field, and may therefore give rise to an anomalous
Hall effect. 
Their coupling to the charge, spin, and orbital degrees of freedom
is particularly transparent in the continuum limit\cite{auerbach,com6}
of (\ref{tb}):
\begin{eqnarray}
H_{\rm cont} &=& \int d^3 r \ [  t_{\rm eff} \psi^{\dagger}
(i\partial_{\mu} +  {e } A_\mu+   a_\mu^s +   a_\mu^n)^2
\psi \ 
+\rho_{\rm eff}^s {\bf z}^\dagger (i \partial_{\mu} +  a^s_{\mu})^2 {\bf z}
+\rho_{\rm eff}^n {\bf w}^\dagger (i \partial_{\mu} +  a^n_{\mu})^2{\bf w}  
+ g {\bf w}^\dagger {\bf Q} (r) {\bf w} ]
\label{cont}
\end{eqnarray}
where $t_{\rm eff}$ is the effective hopping parameter of charge carriers 
and the spin and orbital stiffness 
$\rho_{\rm eff}^{s,n}$ are defined in the sense of a self-consistent 
mean field theory. ${\bf Q}(r)$ is the quenched random
phonon field.
The scale of $\rho_{\rm eff}^{s,n}$ is set by the electron kinetic energy while
the effective hopping of the holes at finite temperatures is
reduced from its bare value $t$ by the fluctuation of the local moments
and the orbital degree of freedom. The $\langle \psi \psi^\dagger \rangle
= 1-x$ factor appearing in the MFT, can be absorbed into resetting 
the boson density 
${\bf z}^\dagger {\bf z} = {\bf w}^\dagger {\bf w} = 1-x$.
We have introduced the electromagnetic gauge field ${\bf A}$
which only couples to the charge of the fermion ($e>0$ since $\psi^\dagger$
creates holes).
The key point in (\ref{cont}) is
that the fermions feel an effective magnetic field
${\bf B}_{\rm eff} = {\bf B} + e^{-1}{\bf b}^s + e^{-1}{\bf b}^n =
\nabla \times ({\bf A} + e^{-1} {\bf a}^s + e^{-1} {\bf a}^n )$
which includes the contribution of the bosonic Berry phases
described by ${\bf a}^{s,n}$.

The spin-orbit Hamiltonian (\ref{so}) includes the coupling to the Berry phase
which is made explicit by substituting
$c_{\sigma ,a} = \psi^\dagger z_\sigma w_a$.
\begin{equation}
H^{\prime}_{\rm so} 
\approx  \int dr^3 \, { u} \langle \psi \psi^{\dagger} \rangle
\, {\bf M } \cdot \nabla \times ({\bf a}^n + {\bf a}^s) 
\end{equation}
where we have assumed that
${\bf z}$ and ${\bf w}$ are slowly varying functions 
and suppressed the terms with derivatives acting on $\psi$.
We observe here that the spin-orbit interaction couples the average
magnetization to the bosonic Berry phases
(topological flux) ${\bf b}^{s,n}$. The latter arises through
non-planar configurations of the ${\bf s} (r)$ and/or ${\bf n}(r)$
via ${\bf b}^s_i (r) = \epsilon_{ijk} {\bf s} \cdot \partial_j {\bf s}
\times \partial_k {\bf s}$ and analogously for ${\bf b}^n$. Note
however that strong Jahn-Teller effect ({\it i.e.} $g \, |{\bf Q}| \gg k_B T$) 
would confine ${\bf n}$ to a plane and suppress ${\bf b}^n$. 
The fact that the spin-orbit interaction couples magnetization to
the topological flux associated with spin (or orbital) textures is
quite general and not restricted to the strong coupling limit
and any slave boson/fermion parametrization.
However, the effects become weak for small $J_{\rm H}/t$.

Let us start
by considering the case of quenched orbital phases, ${\bf b}^n =0$ and
analyse the effect of spin textures in the paramagnetic and ferromagnetic
phases.
In the FM phase ${\bf z}$ bosons are Bose-condensed corresponding
to FM order in ${\bf s}$.
As a result, the spin gauge field ${\bf a}^s$ becomes massive with
the mass proportional to the density of the condensate and thus
there is no uniform contribution to ${\bf b}_s$. On the other
hand in the PM phase we expect to find a quadratic contribution,
$({\bf b}^s)^2$, to the total energy and a finite topological susceptibility
defined as $\chi_T \equiv \partial \langle b^s \rangle / \partial M$.
The calculation proceeds
by integrating out the $\psi$ and ${\bf z}$ fields in Eq.(\ref{cont})
and results in an effective Hamiltonian:
\begin{equation}
H_b  =
\int d^3r \, [ \chi_F  \ ({e } {\bf B} \, + \, {\bf b}^s)^2
+ \ \chi_B  \, ({\bf b}^s)^2 + 
{ u} (1-x) \ {\bf M} \cdot \, {\bf b}^s ] \ 
\label{effb}
\end{equation}
with the diamagnetic susceptibilities
$\chi_F  = \pi^{-2} t_{\rm eff}^2 D(\epsilon_F) $ and
$\chi_B = n(0) \sqrt{2 T \rho^s_{\rm eff}}$ (where
$n(\epsilon) = 1/( e^{(\epsilon - \mu_B)/T} - 1)$
is the occupation number of the ${\bf z}$ bosons) are determined by the
current/current correlators of the fermions and bosons respectively. 
Minimizing with respect to ${\bf b}^s$ yields
\begin{equation}
\langle {\bf b}^s \rangle \,= 
- { u (1-x) \over 2 (\chi_F + \chi_B) } {\bf M} \,- \,
{ e \chi_F \over \chi_F + \chi_B} {\bf B} \ 
\label{solb}
\end{equation}
from which we read off the topological susceptibility
$\chi_T = - {u (1-x) \over 2 (\chi_F + \chi_B) } $. Magnetization here
refers to the average local moment in units of $g_e \mu_B S$ where
$g_e$ is the Lande $g$-factor and $\mu_B$ is the Bohr magneton. 
The topological
flux per plaquette of the lattice is of order $u/\chi_{F,B}$
which we can estimate conservatively by taking $u \sim 1 - 10K$ and
$\chi \sim t$ yielding  $\langle {\bf b}^s \rangle \sim 10^{-3} - 10^{-2} l^{-2}$ where
$l$ is the lattice constant. This flux is comparable to $eB/\hbar c$ for
$1 T$ field.
One also observes that the flux
is maximized for a nearly full band as may be expected for 
the effect arising from interactions.

According to the Ioffe-Larkin phenomenology\cite{ioffe} 
the constraint between the fermionic and the bosonic currents 
which follows from (\ref{tb}) implies that
the Hall resistivity $\rho_{xy}$ is given by
the sum of fermionic $\rho^F_{xy}$ and the bosonic $\rho^B_{xy}$ 
contributions.
Using the semiclassical Hall constants corresponding to fermion
density $n_F = x$ and boson density $n_B = 1-x$, we have
$\rho^F_{xy} = (eB +  b^s)/e^2 n_F$ 
and $\rho^B_{xy} = b^s/e^2 n_B$ so that
\begin{equation}
\rho_{xy} = - {{ u} (1-x) 
(  n_F^{-1}  +   n_B^{-1}  )  
\over 2 e^2 (\chi_F+\chi_B)}\, M
+ {\chi_B n_F^{-1} - \chi_F n_B^{-1} \over \chi_F + \chi_B} {B \over e} \ .
\label{hall}
\end{equation}
from which we identify:
\begin{eqnarray}
R_H &=& {1 \over e} {{\chi_B n_F^{-1} - \chi_F n_B^{-1}} 
\over {\chi_F + \chi_B}}  \cr
R_A &=& - { u (1-x) 
(  n_F^{-1}  +   n_B^{-1}  )  
\over 2 e^2  (\chi_F + \chi_B)} \ 
\label{coeff}
\end{eqnarray}
According to the naive classical estimate, 
${u} > 0$ and $R_A$ has the sign\cite{com10} opposite
to $R_H$. This could explain the sign of the anomalous Hall effect 
observed by Matl {\it et al}\cite{ong}. 
Note also the compensation of $R_H$ by the bosonic
contribution: the large apparent carrier concentration observed in the
experiment suggests that some compensation effect does indeed exist.
However, the calculation of the Hall constant in an interacting system
clearly requires greater effort than the semiclassical estimates.

The above estimate for $\chi_F$ given above is appropriate for the 
metallic paramagnetic phase
(such as observed in the doped rare earth manganites at the values of
$x$ somewhat higher than optimal value for the CMR effect);
at lower dopings the PM phase has very high resistivity, thus we
expect in this situation that $\chi_F \ll \chi_B$ 
($D(\epsilon_F) \rightarrow 0$) and
the relevant carrier density $n_F$ ( $n_F^{-1} \gg n_B^{-1}$)
to be thermally activated.
For the CMR compounds, the insulating PM phase is likely to be 
caused by the large local Jahn-Teller distortions\cite{millis,los} which
split the conduction band into filled orbitally non-degenerate subband
with spectral weight $1-x$ and an empty orbitally degenerate
subband with spectral weight $x$. Conduction is
due to the thermally excited carriers in both subbands and the 
proper determination of the sign and magnitude
of $R_H$ is non-trivial. (Alternatively one may think of
transport in terms of small polarons\cite{jaime}.)
However the appearance, due to spin-orbit interaction,
of the average topological field $\langle {\bf b}^s \rangle$ 
as well as its coupling to
the carriers are robust.
We expect $R_A \approx e^{-1} \chi_T R_H$ with 
$\chi_T = - u(1-x) (2\chi_B)^{-1}$, which follows from (10)
under $\chi_F \ll \chi_B$ and $n_F^{-1} \gg n_B^{-1}$ assumptions but
can be understood simply by saying that 
$\langle {\bf b}^s \rangle$ acts just like magnetic field.

Let us now discuss the ferromagnetic
phase. As already noted, in the FM phase the 
${\bf a}^s$ gauge field acquires a mass due to the divergence of $\chi_B$ at $k=0$. 
This phenomenon is reminiscent of the Meissner effect in a superconductor, 
but there is a crucual difference due to the fact that the topological flux 
on the lattice is not conserved.
A continuum FM spin texture with 
${\bf b}^s_i (r) = \epsilon_{ijk} {\bf s} \cdot \partial_j {\bf s} \times 
\partial_k {\bf s} \neq 0$ is the Belavin-Polyakov\cite{polyakov} skyrmion 
which in a 3D ferromagnet would be a line defect (akin to a flux tube) with 
the line tension $2\pi \rho^s_{\rm eff} |q|$ where $q$ 
is the topological charge equal to the
integral of ${\bf b}^s$ over the crossection of the ``tube" 
divided by $8 \pi$. 
Yet when this continuum solution is transfered to the lattice, 
the quantization of $|q|$ is lost since shrinking the
lateral size of the skyrmion allows the texture to ``fall through" 
the plaquette.
For the same reason (and in contrast to a flux line in a superconductor)
the skyrmion line can have free ends and at $T \neq 0$ there will be 
a finite density of such objects. Alternatively one can argue that the spins 
around any plaquette may be deformed into a texture carrying topological flux 
$q = {l^2 b^s \over 8 \pi}$ at an energy cost
$\alpha \rho^s_{\rm eff} |q|$ ($\alpha$
being a constant of $O(1)$ ) which should be contrasted with
the quadratic energy cost in the PM. 
One can calculate the topological susceptibility from the partition 
function
$Z(M) = \int d{\bf  b}^s \ 
{\rm exp} [ - (\alpha \rho^s_{\rm eff}/8 \pi) |{\bf b}^s|/T - 
 u {\bf b}^s \cdot {\bf M}] $.
One finds $\chi_T \approx - 8 \pi  u T/\alpha \rho^s_{\rm eff}$ for 
$T/\rho^s_{\rm eff} \ll 1$. Since the topological ${\bf b}$
couples to the carriers exactly like the magnetic field we expect:
$R_A = e^{-1} \chi_T R_H$. Note that approaching $T_c$ the 
topological susceptibility $\chi_T$ should be enhanced both
by the softening of
$\rho^s_{\rm eff}$ (due to the reduction of electron kinetic 
energy\cite{millis} which underlies the FM exchange constant) 
and by the contribution of the large scale skyrmion textures which
take advantage of the vanishing spin stiffness at long wavelength.
The enhancement of $\chi_T$ and hence $R_A$ could explain
the observed upturn of $R_A$ below $T_c$.

While the temperature dependence of $R_A$ indicated by the above 
arguments is qualitatively consistent with that observed by 
Matl {\it et al}\cite{ong} 
the region close to $T_c$ requires a more careful treatment. 
At present we neither 
understand why (or whether) $R_A (T) \sim \rho(T)$ nor why both quantities 
peak $20-30K$ $above$ $T_c$. 
An {\it a priory} calculation of $R_H$ is beyond our present scope.
However we did demonstrate how in the presence of interactions the spin-orbit 
coupling can generate direct dependence of $\rho_{xy}$
on magnetization via the appearance of topological flux. 
This should be contrasted to the traditional view\cite{kondo,maranzana} 
of skew-scattering contributing to $\rho_{xy}$ in the $3d$ order via 
$\langle (M-\langle M \rangle)^3\rangle$ which peaks well below $T_c$.

Curiously,
the analysis presented for the PM phase applies equally well to the
description of the orbital Berry phase effects in the orbital ``liquid" state
\cite{nagaosa} ($g|{\bf Q}| \ll k_B T$) where 
the orbital degeneracy is restored
in FM phase. 
While in double exchange problem the Hubbard $U$ played no role as the single 
occupancy was ensured by the Hund's rule splitting, if the orbital degeneracy 
is restored, the effect of $U$ becomes important and necessary for the
appearence of the topological effects. 
In this case, ${\bf b}^s$ in Eq.\ref{effb} and Eq.\ref{solb} is replaced by 
${\bf b}^n$ and the rest of the discussion goes in the same way.
Therefore, if $U$ is large and $g|{\bf Q}|$ small, 
the effect of the orbital Berry phase would provide additional contribution 
to the anomalous Hall coefficient in FM phase.

In conclusion we emphasize the generality of the physical mechanism by which
the topological textures in magnetic systems get polarized by the
magnetization via spin-orbit coupling and contribute to $\rho_{xy}$.
Comparing an itenerant magnet
to the double exchange magnet considered above, the former
corresponds to a weak coupling regime, where
the local moments and hence any possible textures with 
$\langle {\bf S}_i \cdot
{\bf S}_j \times {\bf S}_k \rangle \not= 0$ disappear at $T_c$. 
By contrast, manganites
with their core spins and large $J_H$ are in the strong coupling limit.
While our work
points to the possible origin of the unusual anomalous Hall effect
in doped manganites much further work is required to make a
quantitative comparison.
It would also be interesting to measure the anomalous Hall effect for 
La$_{0.7}$Sr$_{0.3}$MnO$_4$
or the like compound, which remains metallic in the PM phase, which
could be compared with our result on PM metallic phase.
   
We are grateful to N. P. Ong for sharing unpublished data and 
enlightening discussions.
We also thank A. M. Sengupta and S. H. Simon for helpful discussions.
Y.B.K. is supported by NSF grant No. DMR-9632294 and A.J.M. NSF
grant No. DMR-9707701.


\begin{references}
\bibitem{com2} In our model we assume complete alignment of 
the itenerant and the core spin on site so that both the current coupling 
to the local moment and the spin current coupling to local electric field 
take the same form of (\ref{so}).
\bibitem{com6} In general, the spin stiffness $\rho^s_{\rm eff}$
and the orbital stiffness $\rho^n_{\rm eff}$ are different due to
the orbital dependence of the hopping amplitude (non-diagonal in
the orbital space) and the presence of the JT coupling.
\bibitem{com10} If the carrier spin were antialigned with
the local moment, the topological charge would reverse. Further, if
the spin-orbit interaction were dominated by the relativistic term
(spin current/local electric field) the sign of $u$ would also reverse
together with $R_A/R_H$.
\end{references}
\end{document}